\documentclass{emulateapj}

\usepackage{graphicx}
\usepackage{amsmath}
\usepackage{natbib}
\usepackage[usenames]{color}
\usepackage{float}
\usepackage{subfigure}
\usepackage{hyperref}

\newcommand{\galaxy}{\texttt{GALAXY}}
\newcommand{\gyrfalcON}{\texttt{gyrfalcON}}

\newcommand{\refsec}[1]{Section~\ref{#1}}
\newcommand{\reffig}[1]{Fig.~\ref{#1}}

\begin{document}
\title{Black Hole Growth in Disk Galaxies Mediated by the Secular Evolution of Short Bars}
\author{
Min Du\altaffilmark{1, 4}, 
Victor P. Debattista\altaffilmark{2, 3},
Juntai Shen\altaffilmark{1, 4}, 
Luis C. Ho\altaffilmark{5, 6},
Peter Erwin\altaffilmark{7, 8}
}

\altaffiltext{1}{Key Laboratory of Research in Galaxies and Cosmology, Shanghai Astronomical Observatory, Chinese Academy of Sciences, 80 Nandan Road, Shanghai 200030, China, \href{mailto:dumin@shao.ac.cn}{dumin@shao.ac.cn}}
\altaffiltext{2}{Jeremiah Horrocks Institute, University of Central Lancashire, Preston PR1 2HE, UK, \href{mailto:VPDebattista@uclan.ac.uk}{VPDebattista@uclan.ac.uk}}
\altaffiltext{3}{Center for Theoretical Astrophysics and Cosmology, Institute for Computational Science, University of Z\"{u}rich, Winterthurerstrasse 190, CH-8057 Z\"{u}rich, Switzerland}
\altaffiltext{4}{College of Astronomy and Space Sciences, University of Chinese Academy of Sciences, 19A Yuquan Road, Beijing 100049, China, \href{mailto:jshen@shao.ac.cn}{jshen@shao.ac.cn}}
\altaffiltext{5}{Kavli Institute for Astronomy and Astrophysics, Peking University, Beijing 100871, China}
\altaffiltext{6}{Department of Astronomy, School of Physics, Peking University, Beijing 100871, China}
\altaffiltext{7}{Max-Planck-Insitut f\"{u}r extraterrestrische Physik, Giessenbachstrasse, 85748 Garching, Germany}
\altaffiltext{8}{Universit\"{a}ts-Sternwarte M\"{u}nchen, Scheinerstrasse 1, 81679 M\"{u}nchen, Germany}

\begin{abstract}
The growth of black holes (BHs) in disk galaxies lacking classical bulges, which implies an absence of significant mergers, appears to 
be driven by secular processes. Short bars of sub-kiloparsec radius have been hypothesized to be an important mechanism for 
driving gas inflows to small scale, feeding central BHs. In order to quantify the maximum BH mass allowed by this mechanism, we examine 
the robustness of short bars to the dynamical influence of BHs. Large-scale 
bars are expected to be robust, long-lived structures; extremely massive BHs, which are rare, are needed to completely destroy such bars. 
However, we find that short bars, which are generally 
embedded in large-scale outer bars, can be destroyed quickly when BHs of mass $M_{\rm bh}\sim0.05-0.2\%$ of the total stellar mass ($M_\star$) 
are present. In agreement with this prediction, all galaxies observed to host short bars have BHs with a mass fraction less than 
$0.2\%M_\star$. Thus, the dissolution of short inner bars is possible, perhaps even frequent, in the universe. An important 
implication of this result is that inner-bar-driven gas inflows may be terminated when BHs grow to $\sim0.1\%M_\star$. We predict 
that $0.2\%M_\star$ is the maximum mass of BHs allowed if they are fed predominately via inner bars. This value matches 
well the maximum ratio of BH-to-host-galaxy stellar mass observed in galaxies with pseudo-bulges and most 
nearby active galactic nucleus host galaxies. This hypothesis provides a novel explanation for the lower $M_{\rm bh}/M_\star$ in galaxies 
that have avoided significant mergers compared with galaxies with classical bulges.

\end{abstract}



\section{Introduction}
\label{introduction}

It is well established that ellipticals and classical bulges follow tight scaling relations with the masses of their black holes 
\citep[BHs; see the review by][]{Kormendy&Ho13}. Classical bulges are known as remnants of galaxy mergers 
\citep[e.g.][]{Toomre77}; these scaling relations led to the prevalence of models involving merger-driven coevolution of 
BHs and host galaxies \citep[e.g.][]{DiMatteo05, Croton06}. However, bulgeless galaxies and galaxies with pseudo-bulges, 
which have a formation history free of significant mergers, do not follow the same scaling relations as classical bulges and ellipticals 
\citep{Kormendy&Ho13}. Many such disk galaxies hosting supermassive BHs have been found 
\citep[e.g.][]{Filippenko&Ho03, Greene10, Jiang11, DongXiaobo12, JiangNing13, Simmons13, Greene16}, 
suggesting that significant black hole growth can be driven largely by internal secular processes.
Deep observations of the galaxy morphology have showed that BH growth is unlikely to have been driven by significant mergers, at 
least since redshift $z\sim2$ \citep{Gabor09, Georgakakis09, Cisternas11, Schawinski11, Kocevski12, Fan14}. 

Even though bars are considered the most important drivers of secular evolution, how they affect the scaling relations is still not well understood.
Near-infrared surveys show that in the nearby universe about two-thirds of disk galaxies host stellar bars \citep[e.g.][]{Eskridge00, Menendez-Delmestre07}.
Almost one-third of barred galaxies also host a short inner bar of general radius $\lesssim1$ kpc \citep{Erwin&Sparke02, Laine02, Erwin04, Erwin11}, 
such systems are termed double-barred (S2B) galaxies. At sub-kiloparsec scales short inner bars have been hypothesized to be an 
important mechanism for driving gas inflows into the center, efficiently feeding BHs \citep[e.g.][]{Shlosman89, Hopkins10a}. 
Thus, in order to understand the secular growth of BHs, a 
crucial question is under what conditions short inner bars can survive the presence of a BH. 

In triaxial systems, central massive concentrations (CMCs), e.g. BHs, dense stellar clusters, and nuclear disks, can scatter stars 
moving on elongated orbits onto more chaotic ones \citep[e.g.][]{Gerhard&Binney85}. The dissolution of bars under the dynamical 
influence of BHs has been studied via self-consistent $N$-body simulations \citep[e.g.][]{Shen&Sellwood04, Athanassoula05}. 
\citet{Shen&Sellwood04} showed that extremely massive BHs (i.e., more than $4\%$ of the galaxy stellar mass $M_\star$) are necessary to destroy large-scale bars. Measurements 
of BH masses find a mass range $10^{6-10}M_\odot$, and the typical mass ratio of BH-to-host-galaxy stellar mass is $0.1\%$. Thus, 
the BH mass fraction required to destroy bars is rare. Therefore, whether the dissolution of bars 
has ever happened in the universe has been seriously questioned. 

As suggested in \citet{Hozumi12}, shorter bars might be more fragile than normal bars of radius $a_{\rm Bar}=2-4$ kpc. 
In this Letter we constrain how massive a BH can become before destroying inner bars by adding BHs in the S2B models of \citet{Du15}. 
In \refsec{model}, we describe the simulations we use. The destruction of short inner bars by BHs is presented in 
\refsec{simS2B}. A novel explanation of the observed limit in the mass ratio of BH-to-host-galaxy is presented in 
\refsec{MbhMstar}. We summarize our conclusions in \refsec{conclusion}.

\section{Model settings}
\label{model}

All the collisionless models we study involve isolated, initially pure-exponential disks. The simulations are evolved with the 3D cylindrical 
polar option of the \galaxy\ $N$-body code \citep{Sellwood14}. 
The system of units is set to $G=M_{\rm 0}=h_R=1$, where 
$G, M_{\rm 0}$, and $h_R$ are the 
gravitational constant, mass unit, and scale length of the initial disk, respectively. Physical quantities can be obtained by choosing 
appropriate scalings. A reasonable scaling to typical spiral/S0 galaxies is $M_0=4.0\times 10^{10}M_\odot$ and $h_R=2.5$ kpc, 
which gives the unit of time $t_0=\sqrt{h_R^3/GM_0}\simeq9.3$ Myr. 
We use grids measuring $N_R \times N_\phi \times N_z = 58 \times 64 \times375$, which give rise to a force resolution 
of 0.01 in the central regions. The disk consists of 
four million equal-mass particles with softening radius $0.01=25$pc. As the central dynamics are largely dominated by the stellar 
component, we use the same logarithmic rigid halo that \citet{Du15} used to simplify the simulations. 

In this Letter, we examine the robustness of the well-studied short inner bar of the standard S2B model in \citet{Du15, Du16a}, which has 
a disk mass $M_\star=1.5M_0$. The initial Toomre-$Q$ is set to $\sim2.0$ in the outer region; in the inner region, 
Toomre-$Q$ is reduced gradually toward the center reaching a minimum value of 0.5. Thus, the inner bar forms spontaneously from the strong 
bar instability of such a dynamically cool, rotation-dominated, inner disk within a few hundred Myr. 
At steady state, the semi-major axis of the inner bar is $a_{\rm in} \sim 0.3\sim0.75$ kpc. 
The outer bar, which extends to $a_{\rm out}\sim 3.0 \sim 7.5$ kpc, forms slowly in the hotter outer disk.
In this standard S2B model, the inner bar rotates about three times faster than its outer counterpart.

We also examined the clumpy S2B model whose inner disk fragments, forming clumps at the beginning because of using an even colder 
inner disk (minimum Toomre-$Q\sim 0.3$) than the standard S2B. The clumps move toward the center quickly, then coalesce into an inner bar 
\citep[Figure 7 in][]{Du15} that is relatively stronger, i.e., more massive and longer ($a_{\rm in}\sim1.5$ kpc). 

The CMC is introduced as the potential of a rigid Plummer sphere 
\begin{equation}
\centering
      \Phi_{\rm C}(r) = -\frac{GM_{\rm C}(t)}{\sqrt{r^2 + \epsilon^2_{\rm C}}}, 
\end{equation}
where $M_{\rm C}(t)$ and $\epsilon_{\rm C}$ are the mass and softening radius, respectively, of the CMC. As $\epsilon_{\rm C}$ determines the
compactness of the CMC, we use $\epsilon_{\rm C}=0.001=2.5 {\rm pc}$ to mimic a BH. The force from the CMC is added directly 
to each particle from the analytic form, and is therefore independent of the grid resolution. 

In order to mimic the secular growth of BHs, the BH mass is gradually increased after the bars have reached 
a steady state at $t_{\rm C}=300$. The initial CMC mass $M_{\rm C0}=0.0001\%M_\star$ grows smoothly to the maximum mass 
$M_{\rm Cm}$ in $t_{\rm g}=50$ time units, as follows:
\begin{equation}
\centering
      M_{\rm C}(t) = \left\{
      \begin{aligned}
            & M_{\rm C0} & \tau<0 \\
            & (M_{\rm Cm}-M_{\rm C0}) \sin^2(\pi \tau / 2) + M_{\rm C0} & 0 \le \tau \le 1 \\
            & M_{\rm Cm}                       & \tau >1,
      \end{aligned}
      \right.
\label{BHgrowth}
\end{equation}
where $\tau = (t - t_{\rm C}) / t_{\rm g}$. The growth time $t_{\rm g}$ is much 
longer than the dynamical time scale of the central particles. Thus, it can be regarded as an adiabatic growth. The simulations last
$800$ time units, $\simeq7.4$ Gyr in our standard scaling. 

The maximum mass of the BH, $M_{\rm bh}$, is varied in the range of $0.01\%M_\star$ to 
$0.2\%M_\star$. To ensure accurate integration for rapidly moving particles, 
we reduce time steps by half for adding guard shells around the BH \citep{Shen&Sellwood04}. 
The time step is $\Delta t=0.01$ outside of the guard shells. 
We introduce eight shells at $r\le 0.12$, the shortest time step reaches $\Delta t / 2^8$, which is sufficiently small even for the fastest 
moving particles. 

\section{The dissolution of short inner bars}
\label{simS2B}

\subsection{The standard S2B model}
\begin{figure*}[htp]
\centering
        \includegraphics[width=0.3\textwidth, angle=270]{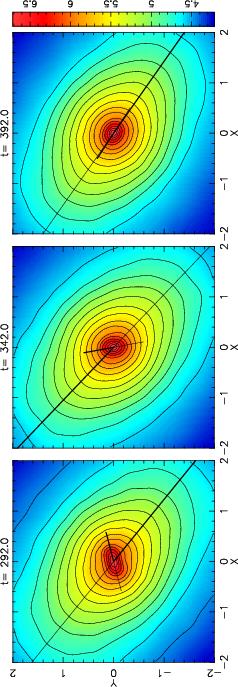}
        \caption{Face-on surface density images of the standard S2B model harboring a BH of mass $M_{\rm bh}=0.2\%M_\star$, showing
            the quick dissolution of the short inner bar within 100 time units ($\sim 1$ Gyr). The density contours in logarithmic space are
            overlaid in black. The long and the short straight lines mark the orientations of the outer and the inner bars, respectively.}
      \label{fig:S2B1}
\end{figure*}

\begin{figure}[htp]
\centering
        \includegraphics[width=0.5\textwidth]{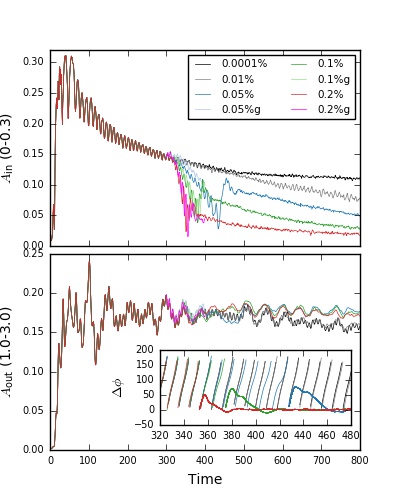}
        \caption{Time evolution of the inner (top) and the outer (bottom) bars of the standard S2B model under the dynamical influence of BHs. 
            The BHs grow to the maximum mass $M_{\rm bh}$ (shown in the legend) from a seed of mass $0.0001\%M_\star$ during $t=300-350$. 
            The cases of $M_{\rm bh}/M_\star=0.0001-0.2\%$ are shown here. $A_{\rm in}$ and $A_{\rm out}$ 
            are defined as the Fourier $m=2$ amplitudes averaged over $R\le 0.3$ and $1.0 \le R \le 3.0$, respectively. The relative position 
            angle of the two bars, $\Delta \phi$, as a function of time is shown in the inset. For cross-check, 
            the results of $M_{\rm bh}/M_\star=0.05-0.2\%$ recalculated using \gyrfalcON\ are labeled as $0.05\%$g, $0.1\%$g, and $0.2\%$g, 
            respectively.}
      \label{fig:S2Bampl}
\end{figure}

As shown in \reffig{fig:S2B1}, the inner bar is completely destroyed within 100 time units by a BH of mass $M_{\rm bh}=0.2\%M_\star\sim10^8M_\odot$. 
\reffig{fig:S2Bampl} shows the evolution of the inner ($A_{\rm in}$) and outer ($A_{\rm out}$) bar amplitudes. 
The deep black profiles represent the model using a tiny BH of fixed mass $0.0001\%M_\star$, in which case the bars are almost unaffected. 
The growth of a $0.05\%M_\star$ BH breaks the dynamical equilibrium of the two decoupled bars within 
a short time. The relative position angle of the two bars, $\Delta \phi$, for all of these models is shown in the inset of \reffig{fig:S2Bampl}. 
Although $A_{\rm in}$ decreases sharply because of the growth of BH, the pattern speed does not change much during the inner bar 
weakening, e.g. $t=300-420$ in the case of $M_{\rm bh}/M_\star=0.05\%$. 
As the inner bar still rotates independently of its outer counterpart, they are still interacting strongly. The evolution of 
$\Delta \phi$ shows that the weakened inner bars may be trapped by the outer bars, i.e., coupling to alignment. The 
coupling process is essentially similar to the case of the aligned S2B 
presented in \citet[][Figure 11]{Du15}: the fast-rotating inner bar slows down sharply before reaching a perpendicular orientation, falling into 
alignment with the outer bar. The key difference from the aligned S2B of that earlier model is that the coupling leaves no prominent face-on
peanut-shaped density contours. The amplitude of the aligned ``inner'' bar decreases slowly in the later stage, until it becomes 
quite round, probably indicating continued orbital scattering. 
In order to verify that the evolution found here is not an artifact of the grids and code use, we rerun the simulations of bar dissolution 
($t=300-400$)
with the momentum-conserving treecode \gyrfalcON\ \citep{Dehnen00, Dehnen14}\footnote{Available at the \texttt{NEMO} repository \citep{Teuben95}: 
\href{url}{http://admit.astro.umd.edu/nemo/}}. Adaptive time steps are used to accurately integrate particle motions close to the BHs. 
\reffig{fig:S2Bampl} includes the results from these tests; in all cases, the results are consistent with those from \galaxy.

Because many massive BHs are already present in the very early universe, likely before disks and bars were established \citep{Fan01, Fan03}, we 
also model this scenario by adding a BH of fixed mass from the beginning of 
the simulations. In order to generate the initial disk for the S2B instability, we use the same initial conditions as the standard S2B model. 
Because of the chaotic nature of galactic disks \citep{Sellwood&Debattista09}, stochasticity can significantly affect  
such models \citep{Du15}. Therefore, we run 10 S2B simulations for each BH mass changing only the random seed for generating the initial particles.
In the case of a pre-existing BH, within $\sim1$ Gyr, the initial disk generally generates an S2B structure similar to that in the standard 
S2B model. With a pre-existing BH of mass $0.2\%M_\star$, we never obtain a steady S2B feature lasting more 
than 1.5 Gyr. In the case of a $0.05\%M_\star$ BH, half the simulations maintain decoupled S2B features for more than 4 Gyr. 

\subsection{Other models}
In the clumpy S2B model, the strong inner bar extends to $1.5$ kpc, while the outer bar is quite weak. Thus, the gravitational torque of 
the outer bar has a negligible effect on the robustness of the inner bar. We regard this model as an extreme case for testing the 
robustness of short bars under the dynamical influence of BH growth.
The inner bar in this case can be destroyed within $2$ Gyr by a BH of mass $0.3\%M_\star$ without coupling between the two bars.
We have examined that a single nuclear bar performs similarly to the inner bar of the clumpy S2B model.
Using a single-barred simulation ($a_{\rm Bar}\sim3-4$ kpc), \citet{Valluri16} investigated the change of the orbital 
families that is induced by the growth of BHs. They showed that BHs of mass $0.2\%M_\star$ can 
destroy most bar-supporting orbits within $R<1.5$ kpc which is consistent with our result. Therefore, $0.2\%M_\star$ is likely to 
be the maximum BH mass allowing the presence of short inner bars in galaxies free of mergers.

\section{The $M_{\rm bh}-M_\star$ relation}
\label{MbhMstar}

\begin{figure}
\centering
      \includegraphics[width=0.45\textwidth]{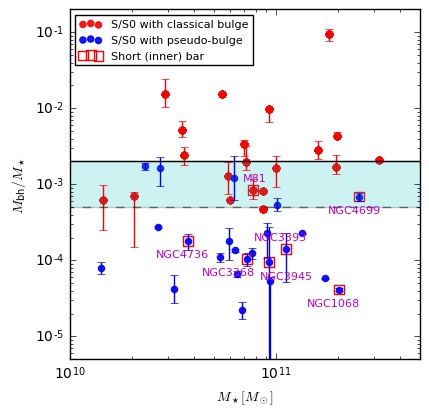}
      \caption{Mass ratios of BH-to-host-galaxy stellar mass summarized in \citet{Kormendy&Ho13}. The BH mass and 
            stellar mass of NGC4699 are adopted from \citet{Saglia16} and \citet{Erwin15}, respectively. The red 
            and blue dots represent the galaxies with classical bulges and pseudo-bulges, respectively. Their BH masses are all 
            measured via stellar dynamics. The error comes from the uncertainty in $M_{\rm bh}$ measurement. The maximum BH mass fraction obtained 
            from our simulations is overlaid in the cyan shaded region. The galaxies having a short (inner) bar are marked with red squares.} 
      \label{BHvsGalKH13}
\end{figure}

\begin{figure*}
\centering
      \includegraphics[width=0.9\textwidth]{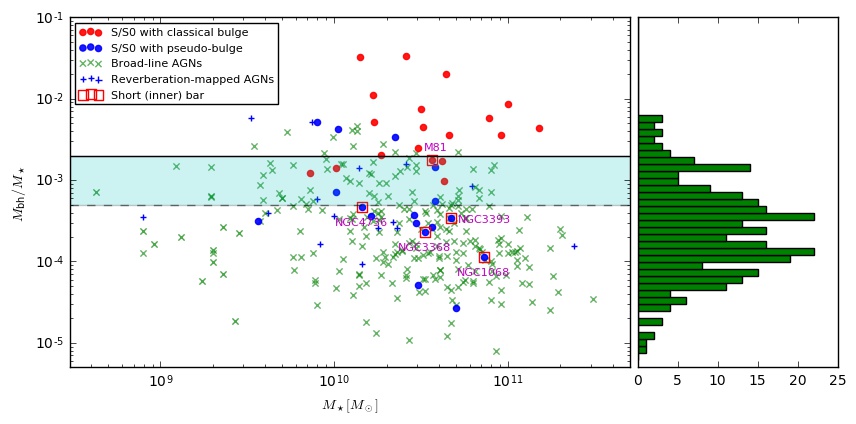}
      \caption{Mass ratios of BH-to-host-galaxy stellar mass estimated by \citet{Reines&Volonteri15}. 
            The BH masses of 256 broad-line AGNs (green crosses, including 12 dwarf galaxies) and 15 reverberation-mapped AGNs 
            (blue plusses) are adopted from \citet{Reines&Volonteri15} and \citet{Bentz&Katz15}, respectively. The blue
            and red dots represent the objects from \citet{Kormendy&Ho13} shown above in \reffig{BHvsGalKH13}, BH masses for which are 
            measured by stellar dynamics. Stellar masses are estimated using the same method as the AGN host galaxies to avoid 
            systematic deviation. The red squares mark the galaxies having short (inner) bars. NGC3945 and NGC4699 are not included 
            in \citet{Reines&Volonteri15}. In the right panel, the histogram shows the number distribution of mass ratios of the AGN 
            host galaxies and galaxies with pseudo-bulges.} 
      \label{BHvsGal}
\end{figure*}

If short bars are the primary drivers of gas inflows feeding BHs, the dissolution of short bars should halt the secular growth 
of BHs. Thus, the mass of BHs may stall at $\sim 0.1\%M_\star$. In observations, $M_\star$ 
is generally approximated by assuming reasonable mass-to-light 
($M_\star/L$) ratios. Many short bars may have been destroyed after they played an important role in feeding BHs.
After short bars are destroyed, continuing star formation in spiral galaxies will result in an even lower mass ratio of 
BH-to-host-galaxy. Our simulations therefore constrain the maximum $M_{\rm bh}/M_\star$ allowed by secular evolution. 

In Figs. \ref{BHvsGalKH13} and \ref{BHvsGal}, we show different measurements of $M_{\rm bh}/M_\star$ from \citet{Kormendy&Ho13} and 
\citet{Reines&Volonteri15}, respectively. The criterion obtained from the simulations is overlaid as the shaded region. 
In \reffig{BHvsGalKH13}, except for \object{NGC4699}, all the data are taken from \citet{Kormendy&Ho13} where the BH masses are measured 
using dynamical methods and the stellar masses are based on the $M_\star/L$ of \citet{Bell03}.
These spiral/S0 galaxies are classified into two groups based on the morphology of their bulges. 
According to the decomposition of \citet{Erwin15}, \object{NGC4699} is a galaxy with a composite bulge containing a primary pseudo-bulge (mass 
$\sim 36\%M_\star$) and a small classical bulge ($\sim11\%M_\star$). Thus, \object{NGC4699} should also be classified as a galaxy with 
a pseudo-bulge. Disk galaxies with pseudo-bulges tend to fall below the mass ratio of 
BH-to-host-galaxy found in galaxies with classical bulges and elliptical galaxies \citep{Reines&Volonteri15}. Such a clear separation 
suggests that the BHs evolve differently. Exponential, disky pseudo-bulges are generally expected to be generated by internal secular evolution 
without significant mergers. Therefore, the growth of BHs in galaxies with pseudo-bulges and bulgeless galaxies should be 
dominated by secular processes, including inner bar-driven gas inflows. 

Mergers are rare in the local universe. Most active galactic nuclei (AGNs) may also be triggered by secular processes. 
Thus, the BH masses in AGN host galaxies are likely to be lower than the 
maximum BH mass allowed by secular evolution. In \reffig{BHvsGal}, we include the measurements of 256 broad-line AGNs from 
\citet{Reines&Volonteri15} and 15 reverberation-mapped AGNs from \citet{Bentz&Katz15}. The stellar masses in \reffig{BHvsGalKH13} 
are systematically higher by a factor of $\sim2$ than those in \reffig{BHvsGal}, which were estimated by the color-dependent $M_\star/L$ presented 
by \citet{Zibetti09}. 
The histogram in the right panel of \reffig{BHvsGal} shows the number distribution of the mass ratio of the AGN host galaxies and 
the galaxies with pseudo-bulges. Less galaxies are located at the cyan shaded region; mass ratios larger than $0.002$ are rare. Thus, 
maximum BH mass suggested by our simulations is roughly consistent with the observations. It suggests that
BH growth via secular evolution is terminated by the dissolution of short bars.

Short inner bars are expected to be gradually destroyed once $M_{\rm bh}\sim0.05-0.2\%M_\star$. Five S2Bs 
\citep[\object{NGC1068, NGC3368, NGC3393, NGC3945, and NGC4736}; see][]{Erwin04} are marked by red squares in 
Figs. \ref{BHvsGalKH13} and \ref{BHvsGal}. All of them have lower
mass ratios than the upper boundary of the cyan region. In \object{M81}, a short inner bar may be embedded in a weak outer bar 
\citep{Gutierrez11}, although it is generally considered to have a significant classical bulge. \object{NGC4699} only harbors a single short bar.  
The absence of strong outer bars allows short bars to persist for longer because of the lack of interaction 
between two bars. Thus, it is not surprising that \object{M81} and \object{NGC4699} are present at the cyan shaded region. 
However, no S2Bs are above this range.

\section{Conclusions}
\label{conclusion}

BH growth in disk galaxies without significant mergers is expected to be driven by secular processes. Sub-kiloparsec-scale short bars, 
generally embedded in large-scale bars, have been hypothesized to be an important mechanism for driving gas inflows, feeding BHs. 
Whether and under what conditions short bars can survive the presence of a BH can provide a crucial test of this scenario.
By adding BHs at the center of short-bar simulations, we have
found that short bars are likely to be destroyed by BHs with masses at
least $0.05-0.2\%$ of the galaxy stellar mass; mass ratios larger than 
$0.2\% M_{\star}$ always lead to destruction of short bars. This maximum 
mass ratio coincides well with the observed upper limit on BH mass 
ratios for real disk galaxies. Therefore, we provide a possible 
explanation for the lower $M_{\rm bh}/M_\star$ in disk galaxies: the 
BH growth via secular evolution may be terminated by the dissolution 
of short inner bars. 

We sketch a potential evolutionary path of the central structures of disk galaxies as follows. In gas-rich galaxies, gas is 
funneled into central regions by large-scale galactic bars, gradually building a stellar nuclear disk. Short inner bars form spontaneously in 
nuclear disks after sufficient accumulation of dynamically cold stellar material \citep{Du15}. Then, short inner bars funnel gas further into 
the center, feeding BHs. According to the criterion in this Letter, inner bars are destroyed when the BH mass grows to 
$\sim0.1\%M_\star$. The dissolution of inner bars in turn slows down, or even stops, the growth of BHs. 
Thus, BH growth in disk galaxies is mediated by the secular evolution of short bars in the absence of mergers.

However, previous observations have not found a clear relation between inner bars and AGNs \citep[e.g.][]{Regan&Mulchaey99, Martini01, Erwin&Sparke02, Laine02}. 
One possible reason is that catching BH fueling in the act is challenging, as one AGN activity episode typically lasts just a few million years. 
Another possible reason is that AGN fueling may be dominated by some unknown stochastic processes very local to the BH, as 
the radius that inner-bar-driven gas inflows reach may still be too far away from the BH's accretion disk.
Or the fraction of short inner bars may be underestimated because of dust obscuration, particularly in late-type galaxies. 
Furthermore, other non-axisymmetric nuclear structures, e.g. spirals, may also drive gas inflows feeding BHs \citep{Hopkins10}. Thus, 
AGNs may not be completely stalled after inner bars are destroyed. Therefore, it is perhaps not surprising that no clear relation is reported. The 
previous studies cannot rule out short bars as a potentially important mechanism of promoting BH growth in disk galaxies. Whether or not other 
drivers of gas inflows are also suppressed by BH growth will be studied in future work. 

\begin{acknowledgments}
M.D. thanks the Jeremiah Horrocks Institute of the University of Central Lancashire for their hospitality during a three-month visit
while this Letter was in progress. 
We acknowledge support from an {\it Newton Advanced Fellowship} awarded by the Royal Society and 
the Newton Fund, and from the CAS/SAFEA International Partnership Program for Creative Research Teams. 
The research presented here is partially supported by the 973 Program of China under grant No. 2014CB845700, 
by the National Natural Science Foundation of China under grant Nos.11333003, 11322326, and by a China-Chile joint grant from CASSACA. 
This work made use of the facilities of the Center for High Performance 
Computing at Shanghai Astronomical Observatory. V.P.D. is supported by STFC Consolidated grant ST/M000877/1 and acknowledges the 
personal support of George Lake, and of the Pauli Center for Theoretical Studies, which is supported by the Swiss National Science 
Foundation (SNF), the University of Z\"urich, and ETH Z\"urich. The work of L.C.H. was supported by National Key Program for 
Science and Technology Research and Development grant 2016YFA0400702.
\end{acknowledgments}

\bibliographystyle{apj}

\begin{thebibliography}{}
\expandafter\ifx\csname natexlab\endcsname\relax\def\natexlab#1{#1}\fi

\bibitem[{{Athanassoula} {et~al.}(2005){Athanassoula}, {Lambert}, \&
  {Dehnen}}]{Athanassoula05}
{Athanassoula}, E., {Lambert}, J.~C., \& {Dehnen}, W. 2005, \mnras, 363, 496


\bibitem[{{Bell} {et~al.}(2003){Bell}, {McIntosh}, {Katz}, \&
  {Weinberg}}]{Bell03}
{Bell}, E.~F., {McIntosh}, D.~H., {Katz}, N., \& {Weinberg}, M.~D. 2003, \apjs,
  149, 289

\bibitem[{{Bentz} \& {Katz}(2015)}]{Bentz&Katz15}
{Bentz}, M.~C., \& {Katz}, S. 2015, \pasp, 127, 67

\bibitem[{{Cisternas} {et~al.}(2011){Cisternas}, {Jahnke}, {Inskip},
  {Kartaltepe}, {Koekemoer}, {Lisker}, {Robaina}, {Scodeggio}, {Sheth},
  {Trump}, {Andrae}, {Miyaji}, {Lusso}, {Brusa}, {Capak}, {Cappelluti},
  {Civano}, {Ilbert}, {Impey}, {Leauthaud}, {Lilly}, {Salvato}, {Scoville}, \&
  {Taniguchi}}]{Cisternas11}
{Cisternas}, M., {Jahnke}, K., {Inskip}, K.~J., {et~al.} 2011, \apj, 726, 57

\bibitem[{{Croton} {et~al.}(2006){Croton}, {Springel}, {White}, {De Lucia},
  {Frenk}, {Gao}, {Jenkins}, {Kauffmann}, {Navarro}, \& {Yoshida}}]{Croton06}
{Croton}, D.~J., {Springel}, V., {White}, S.~D.~M., {et~al.} 2006, \mnras, 365,
  11

\bibitem[Dehnen(2000)]{Dehnen00} Dehnen, W.\ 2000, \apjl, 536, L39 

\bibitem[Dehnen(2014)]{Dehnen14} Dehnen, W.\ 2014, Astrophysics Source Code Library, ascl:1402.031 

\bibitem[{{Di Matteo} {et~al.}(2005){Di Matteo}, {Springel}, \&
  {Hernquist}}]{DiMatteo05}
{Di Matteo}, T., {Springel}, V., \& {Hernquist}, L. 2005, \nat, 433, 604

\bibitem[{{Dong} {et~al.}(2012){Dong}, {Ho}, {Yuan}, {Wang}, {Fan}, {Zhou}, \&
  {Jiang}}]{DongXiaobo12}
{Dong}, X.-B., {Ho}, L.~C., {Yuan}, W., {et~al.} 2012, \apj, 755, 167

\bibitem[{{Du} {et~al.}(2016){Du}, {Debattista}, {Shen}, \&
  {Cappellari}}]{Du16a}
{Du}, M., {Debattista}, V.~P., {Shen}, J., \& {Cappellari}, M. 2016, \apj, 828,
  14

\bibitem[{{Du} {et~al.}(2015){Du}, {Shen}, \& {Debattista}}]{Du15}
{Du}, M., {Shen}, J., \& {Debattista}, V.~P. 2015, \apj, 804, 139

\bibitem[{{Erwin}(2004)}]{Erwin04}
{Erwin}, P. 2004, \aap, 415, 941

\bibitem[{{Erwin}(2011)}]{Erwin11}
---. 2011, Memorie della Societa Astronomica Italiana Supplementi, 18, 145

\bibitem[{{Erwin} \& {Sparke}(2002)}]{Erwin&Sparke02}
{Erwin}, P., \& {Sparke}, L.~S. 2002, \aj, 124, 65

\bibitem[{{Erwin} {et~al.}(2015){Erwin}, {Saglia}, {Fabricius}, {Thomas},
  {Nowak}, {Rusli}, {Bender}, {Vega Beltr{\'a}n}, \& {Beckman}}]{Erwin15}
{Erwin}, P., {Saglia}, R.~P., {Fabricius}, M., {et~al.} 2015, \mnras, 446, 4039

\bibitem[{{Eskridge} {et~al.}(2000){Eskridge}, {Frogel}, {Pogge}, {Quillen},
  {Davies}, {DePoy}, {Houdashelt}, {Kuchinski}, {Ram{\'{\i}}rez}, {Sellgren},
  {Terndrup}, \& {Tiede}}]{Eskridge00}
{Eskridge}, P.~B., {Frogel}, J.~A., {Pogge}, R.~W., {et~al.} 2000, \aj, 119,
  536

\bibitem[{{Fan} {et~al.}(2014){Fan}, {Fang}, {Chen}, {Li}, {Lv}, {Knudsen}, \&
  {Kong}}]{Fan14}
{Fan}, L., {Fang}, G., {Chen}, Y., {et~al.} 2014, \apjl, 784, L9

\bibitem[{{Fan} {et~al.}(2001){Fan}, {Narayanan}, {Lupton}, {Strauss}, {Knapp},
  {Becker}, {White}, {Pentericci}, {Leggett}, {Haiman}, {Gunn}, {Ivezi{\'c}},
  {Schneider}, {Anderson}, {Brinkmann}, {Bahcall}, {Connolly}, {Csabai}, {Doi},
  {Fukugita}, {Geballe}, {Grebel}, {Harbeck}, {Hennessy}, {Lamb}, {Miknaitis},
  {Munn}, {Nichol}, {Okamura}, {Pier}, {Prada}, {Richards}, {Szalay}, \&
  {York}}]{Fan01}
{Fan}, X., {Narayanan}, V.~K., {Lupton}, R.~H., {et~al.} 2001, \aj, 122, 2833

\bibitem[{{Fan} {et~al.}(2003){Fan}, {Strauss}, {Schneider}, {Becker}, {White},
  {Haiman}, {Gregg}, {Pentericci}, {Grebel}, {Narayanan}, {Loh}, {Richards},
  {Gunn}, {Lupton}, {Knapp}, {Ivezi{\'c}}, {Brandt}, {Collinge}, {Hao},
  {Harbeck}, {Prada}, {Schaye}, {Strateva}, {Zakamska}, {Anderson},
  {Brinkmann}, {Bahcall}, {Lamb}, {Okamura}, {Szalay}, \& {York}}]{Fan03}
{Fan}, X., {Strauss}, M.~A., {Schneider}, D.~P., {et~al.} 2003, \aj, 125, 1649

\bibitem[{{Filippenko} \& {Ho}(2003)}]{Filippenko&Ho03}
{Filippenko}, A.~V., \& {Ho}, L.~C. 2003, \apjl, 588, L13

\bibitem[{{Gabor} {et~al.}(2009){Gabor}, {Impey}, {Jahnke}, {Simmons}, {Trump},
  {Koekemoer}, {Brusa}, {Cappelluti}, {Schinnerer}, {Smol{\v c}i{\'c}},
  {Salvato}, {Rhodes}, {Mobasher}, {Capak}, {Massey}, {Leauthaud}, \&
  {Scoville}}]{Gabor09}
{Gabor}, J.~M., {Impey}, C.~D., {Jahnke}, K., {et~al.} 2009, \apj, 691, 705

\bibitem[{{Georgakakis} {et~al.}(2009){Georgakakis}, {Coil}, {Laird},
  {Griffith}, {Nandra}, {Lotz}, {Pierce}, {Cooper}, {Newman}, \&
  {Koekemoer}}]{Georgakakis09}
{Georgakakis}, A., {Coil}, A.~L., {Laird}, E.~S., {et~al.} 2009, \mnras, 397,
  623

\bibitem[{{Gerhard} \& {Binney}(1985)}]{Gerhard&Binney85}
{Gerhard}, O.~E., \& {Binney}, J. 1985, \mnras, 216, 467

\bibitem[{{Greene} {et~al.}(2010){Greene}, {Peng}, {Kim}, {Kuo}, {Braatz},
  {Impellizzeri}, {Condon}, {Lo}, {Henkel}, \& {Reid}}]{Greene10}
{Greene}, J.~E., {Peng}, C.~Y., {Kim}, M., {et~al.} 2010, \apj, 721, 26

\bibitem[{{Greene} {et~al.}(2016){Greene}, {Seth}, {Kim}, {L{\"a}sker},
  {Goulding}, {Gao}, {Braatz}, {Henkel}, {Condon}, {Lo}, \& {Zhao}}]{Greene16}
{Greene}, J.~E., {Seth}, A., {Kim}, M., {et~al.} 2016, \apjl, 826, L32

\bibitem[{{Guti{\'e}rrez} {et~al.}(2011){Guti{\'e}rrez}, {Erwin}, {Aladro}, \&
  {Beckman}}]{Gutierrez11}
{Guti{\'e}rrez}, L., {Erwin}, P., {Aladro}, R., \& {Beckman}, J.~E. 2011, \aj,
  142, 145

\bibitem[{{Hopkins} \& {Quataert}(2010)}]{Hopkins10a}
{Hopkins}, P.~F., \& {Quataert}, E. 2010, \mnras, 407, 1529

\bibitem[{{Hopkins} {et~al.}(2010){Hopkins}, {Bundy}, {Croton}, {Hernquist},
  {Keres}, {Khochfar}, {Stewart}, {Wetzel}, \& {Younger}}]{Hopkins10}
{Hopkins}, P.~F., {Bundy}, K., {Croton}, D., {et~al.} 2010, \apj, 715, 202

\bibitem[Hozumi(2012)]{Hozumi12} Hozumi, S.\ 2012, \pasj, 64, 5 

\bibitem[{{Jiang} {et~al.}(2013){Jiang}, {Ho}, {Dong}, {Yang}, \&
  {Wang}}]{JiangNing13}
{Jiang}, N., {Ho}, L.~C., {Dong}, X.-B., {Yang}, H., \& {Wang}, J. 2013, \apj,
  770, 3

\bibitem[{{Jiang} {et~al.}(2011){Jiang}, {Greene}, {Ho}, {Xiao}, \&
  {Barth}}]{Jiang11}
{Jiang}, Y.-F., {Greene}, J.~E., {Ho}, L.~C., {Xiao}, T., \& {Barth}, A.~J.
  2011, \apj, 742, 68

\bibitem[{{Kocevski} {et~al.}(2012){Kocevski}, {Faber}, {Mozena}, {Koekemoer},
  {Nandra}, {Rangel}, {Laird}, {Brusa}, {Wuyts}, {Trump}, {Koo}, {Somerville},
  {Bell}, {Lotz}, {Alexander}, {Bournaud}, {Conselice}, {Dahlen}, {Dekel},
  {Donley}, {Dunlop}, {Finoguenov}, {Georgakakis}, {Giavalisco}, {Guo},
  {Grogin}, {Hathi}, {Juneau}, {Kartaltepe}, {Lucas}, {McGrath}, {McIntosh},
 {Mobasher}, {Robaina}, {Rosario}, {Straughn}, {van der Wel}, \&
  {Villforth}}]{Kocevski12}
{Kocevski}, D.~D., {Faber}, S.~M., {Mozena}, M., {et~al.} 2012, \apj, 744, 148

\bibitem[{{Kormendy} \& {Ho}(2013)}]{Kormendy&Ho13}
{Kormendy}, J., \& {Ho}, L.~C. 2013, \araa, 51, 511

\bibitem[{{Laine} {et~al.}(2002){Laine}, {Shlosman}, {Knapen}, \&
  {Peletier}}]{Laine02}
{Laine}, S., {Shlosman}, I., {Knapen}, J.~H., \& {Peletier}, R.~F. 2002, \apj,
  567, 97

\bibitem[{{Martini} {et~al.}(2001){Martini}, {Pogge}, {Ravindranath}, \&
  {An}}]{Martini01}
{Martini}, P., {Pogge}, R.~W., {Ravindranath}, S., \& {An}, J.~H. 2001, \apj,
  562, 139

\bibitem[Men{\'e}ndez-Delmestre et al.(2007)]{Menendez-Delmestre07} 
Men{\'e}ndez-Delmestre, K., Sheth, K., Schinnerer, E., Jarrett, T.~H., \& Scoville, N.~Z.\ 2007, \apj, 657, 790 

\bibitem[{{Regan} \& {Mulchaey}(1999)}]{Regan&Mulchaey99}
{Regan}, M.~W., \& {Mulchaey}, J.~S. 1999, \aj, 117, 2676

\bibitem[{{Reines} \& {Volonteri}(2015)}]{Reines&Volonteri15}
{Reines}, A.~E., \& {Volonteri}, M. 2015, \apj, 813, 82

\bibitem[{{Saglia} {et~al.}(2016){Saglia}, {Opitsch}, {Erwin}, {Thomas},
  {Beifiori}, {Fabricius}, {Mazzalay}, {Nowak}, {Rusli}, \&
  {Bender}}]{Saglia16}
{Saglia}, R.~P., {Opitsch}, M., {Erwin}, P., {et~al.} 2016, \apj, 818, 47

\bibitem[{{Schawinski} {et~al.}(2011){Schawinski}, {Treister}, {Urry},
  {Cardamone}, {Simmons}, \& {Yi}}]{Schawinski11}
{Schawinski}, K., {Treister}, E., {Urry}, C.~M., {et~al.} 2011, \apjl, 727, L31

\bibitem[{{Sellwood}(2014)}]{Sellwood14}
{Sellwood}, J.~A. 2014, ArXiv e-prints, arXiv:1406.6606

\bibitem[{{Sellwood} \& {Debattista}(2009)}]{Sellwood&Debattista09}
{Sellwood}, J.~A., \& {Debattista}, V.~P. 2009, \mnras, 398, 1279

\bibitem[{{Shen} \& {Sellwood}(2004)}]{Shen&Sellwood04}
{Shen}, J., \& {Sellwood}, J.~A. 2004, \apj, 604, 614

\bibitem[{{Shlosman} {et~al.}(1989){Shlosman}, {Frank}, \&
  {Begelman}}]{Shlosman89}
{Shlosman}, I., {Frank}, J., \& {Begelman}, M.~C. 1989, \nat, 338, 45

\bibitem[{{Simmons} {et~al.}(2013){Simmons}, {Lintott}, {Schawinski}, {Moran},
  {Han}, {Kaviraj}, {Masters}, {Urry}, {Willett}, {Bamford}, \&
  {Nichol}}]{Simmons13}
{Simmons}, B.~D., {Lintott}, C., {Schawinski}, K., {et~al.} 2013, \mnras, 429,
  2199

\bibitem[Teuben(1995)]{Teuben95} Teuben, P.\ 1995, Astronomical Data Analysis Software and Systems IV, 77, 398 

\bibitem[{{Toomre}(1977)}]{Toomre77}
{Toomre}, A. 1977, in Evolution of Galaxies and Stellar Populations, ed. B.~M.
  {Tinsley} \& R.~B.~G. {Larson}, D.~Campbell, 401

\bibitem[{{Valluri} {et~al.}(2016){Valluri}, {Shen}, {Abbott}, \&
  {Debattista}}]{Valluri16}
{Valluri}, M., {Shen}, J., {Abbott}, C., \& {Debattista}, V.~P. 2016, \apj,
  818, 141

\bibitem[{{Zibetti} {et~al.}(2009){Zibetti}, {Charlot}, \& {Rix}}]{Zibetti09}
{Zibetti}, S., {Charlot}, S., \& {Rix}, H.-W. 2009, \mnras, 400, 1181

\end{thebibliography}

\end{document}